\documentclass[12pt]{article}

\usepackage{amssymb,amsmath}
\usepackage{emlines2}

\newcommand{\tr}{\mbox{ Tr }}

\begin{document}

\title{$~$
Parquet Approximation \\
for Large $N$ Matrix Higgs Model
$~$\\
$~$\\}

\author{
A.~Shishanin and
I.~Ziyatdinov\footnote{\texttt{isk@ziyatdin.phys.msu.su}}
\vspace{4mm}\\
Department of Physics, Moscow State University, \\
Moscow, 119899, Russia \vspace{1mm}
\\
}

\date {$~$}
\maketitle

\begin {abstract}
The parquet approximation in the matrix Higgs model is considered.
We demonstrate analytically that in the large $N$ limit the
parquet approximation gives an satisfying agreement with the exact
results. It is shown that the parquet planar series can be derived
by means of the generating functional.
\end {abstract}

\section{Introduction}

The parquet approximation was proposed by Landau, Abrikosov and
Khalatnikov in order to develop a self-consistent method of
studying the non-perturbative domain in quantum electrodynamics
\cite{LAK}. Later on, this approximation has been used for various
models of quantum field theory \cite{TM}. The parquet
approximation leads to a closed system of integro-differential
equations which have meaning not only for small but also for large
values of coupling constant. The main problem that prevents direct
application of the parquet approximation to an arbitrary gauge
theory is that the parquet approximation violates the gauge
invariance of the theory.

Another non-perturbative approach used in the quantum field theory
is the planar approximation, a method of studying theories like
$SU$($N$) QCD in the large $N$ limit~\cite{tH}. It enables us to
understand most specific features of QCD~\cite{Witten}, and unlike the
parquet approach it does not break the gauge invariance. But to
perform analytical investigations one has to calculate the sum of
all planar diagrams, and this problem is solved only for a few
simple models such as zero- and one-dimensional matrix model
and two-dimensional QCD~\cite{tH},~\cite{BIPZ}.

There is a hope to construct a method that will combine advantages
of both planar and parquet approaches.

In \cite{Ar96} the planar parquet approximation was defined and
applied to zero-dimensional matrix models with cubic and quartic
interaction terms. It was demonstrated that the planar parquet
approximation gives an excellent agreement with the exact results.
The aim of this paper is to find out whether this approximation
can give any sensible results in the matrix Higgs model. In order
to answer this question we calculate Green functions within both
approaches and compare them.

The paper is organized as follows. In Section~2 we define the
planar parquet approximation and use it for studying the matrix
Higgs model. In~Section~3 the same model is considered by means of
the steepest descent method. In~Section~4 the generating
functional for the whole parquet planar series is constructed. It
is shown that this functional can be regarded as a restricted
planar one.

\section{Planar parquet approximation} \label{section1}

In this section we define and investigate the planar parquet
approximation for the zero-dimensional matrix Higgs model. To
introduce it carefully it is useful to recall what is already
known.

The parquet planar approximation for the
$d$-dimensional matrix model with cubic and quartic interaction
has been defined in \cite{Ar96}. Let us consider the matrix model
with the action
\begin{equation}
    S=\int d^dx \tr\left(\frac 1 2\partial M\cdot \partial M+\frac 1 2
    m^2\,M^2-\frac{\lambda}{3\sqrt N}M^3+\frac{g}{4N}M^4\right),
\end{equation}
here $M$ is a hermitian matrix $N\times N$.

The planar Green functions
\begin{equation}\label{Greens}
    \Pi_n(x_1,...,x_n)=\lim_{N \to
    \infty}\frac{1}{N^{1+n/2}}\dfrac{\int DM \tr( M(x_1)... M(x_n))\exp
    (-S)}{\int DM \exp (-S)}
\end{equation}
satisfy the planar Schwinger--Dyson equations \cite{Haan,Ar81}
\begin{multline}\label{Schwinger-Dyson}
    (-\bigtriangleup  +m^2)_{x_1}\Pi_{n}(x_{1},...,x_{n})\\
    -\lambda\Pi_{n+1}(x_{1},x_{1},x_{2},...,x_{n})+g
    \Pi_{n+2}(x_{1},x_{1},x_{1},x_{2},...,x_{n})=\\
    =\sum _{m=2}^{n}\delta (x_{1}-x_{m})\Pi_{m-2}(x_{2},...,x_{m-1})
    \Pi_{n-m}(x_{m+1},...,x_{n})=0.
\end{multline}
This set of equations is infinite. So the purpose of the planar
parquet approximation is to find conditions when
(\ref{Schwinger-Dyson}) becomes closed, i.e. the planar parquet
approximation is defined as an approximative perturbative solution
of (\ref{Schwinger-Dyson}) that takes into account only a part of
the full series of coupling constant.

It can be done by using a so-called skeleton expansion which
contains only a subset of all planar diagrams. More precisely, the
diagrams from this subset contains no bare propagators, three- and
four-vertices insertions. The basic Green functions, i.e. the
\mbox{two-,} three- and four-point functions are defined as
solutions of a set of integro-differential equations. The
zero-dimensional case reduces it to the set of algebraic
equations. One can represent them graphically

\vspace{5mm}

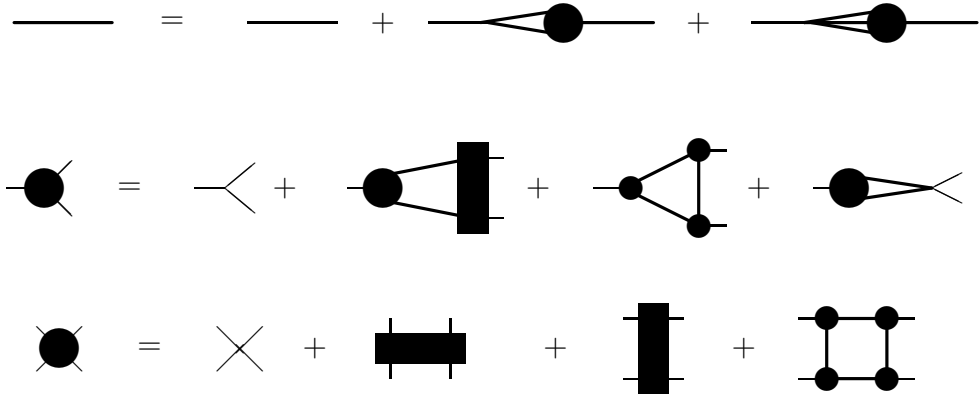
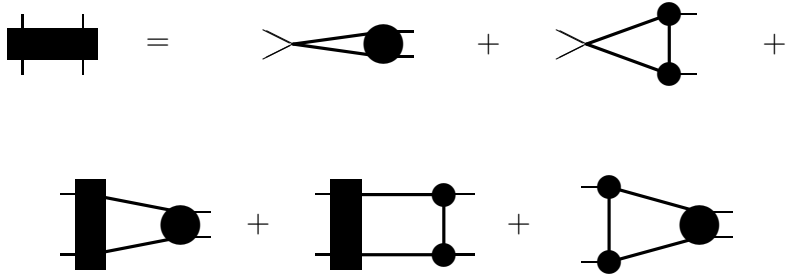
\unitlength=1.00mm \special{em:linewidth 0.4pt}
\linethickness{0.4pt}
\begin{picture}(145.00,17.60)(18,0)
\special{em:linewidth 0.4pt} \linethickness{0.4pt}
\put(38.00,15.00){\makebox(0,0)[cc]{$=$}}
\put(60.00,15.00){\line(-1,0){12.00}}
\put(133.00,15.00){\circle*{5.20}}
\put(115.00,15.00){\line(1,0){8.00}}
\put(90.00,15.00){\circle*{5.20}}
\put(72.00,15.00){\line(1,0){8.00}}
\put(66.00,15.00){\makebox(0,0)[cc]{$+$}}
\put(108.00,15.00){\makebox(0,0)[cc]{$+$}} \special{em:linewidth
1.2pt} \linethickness{1.2pt}
\emline{122.00}{15.00}{1}{145.00}{15.00}{2}
\emline{122.00}{15.00}{3}{133.00}{16.67}{4}
\emline{133.00}{13.33}{5}{123.00}{15.00}{6}
\emline{30.00}{15.00}{7}{17.00}{15.00}{8}
\emline{79.00}{15.00}{9}{90.00}{16.67}{10}
\emline{90.00}{13.33}{11}{80.00}{15.00}{12}
\emline{90.00}{15.00}{13}{102.00}{15.00}{14}
\end{picture}

\unitlength=1.00mm \special{em:linewidth 0.4pt}
\linethickness{0.4pt}
\begin{picture}(142.00,21.67)(15,0)
\special{em:linewidth 0.4pt} \linethickness{0.4pt}
\put(63.00,13.33){\line(0,1){3.33}}
\put(18.00,15.00){\circle*{5.20}}
\put(13.00,15.00){\line(1,0){5.00}}
\put(21.67,18.67){\line(-1,-1){3.67}}
\put(18.00,15.00){\line(1,-1){3.67}}
\put(105.00,20.00){\circle*{3.33}}
\put(105.00,10.00){\circle*{3.33}}
\put(105.00,20.00){\line(1,0){3.67}}
\put(105.00,10.00){\line(1,0){3.67}}
\put(95.00,15.00){\line(-1,0){4.00}}
\put(29.33,15.00){\makebox(0,0)[cc]{$=$}}
\put(42.00,15.00){\line(6,5){4.00}}
\put(42.00,15.00){\line(6,-5){4.00}}
\put(38.00,15.00){\line(1,0){4.00}}
\put(63.00,15.00){\circle*{5.20}}
\put(63.00,15.00){\line(-1,0){4.67}}
\put(75.00,19.00){\line(1,0){4.00}}
\put(75.00,11.00){\line(1,0){4.00}}
\put(73.00,9.00){\rule{4.00\unitlength}{12.00\unitlength}}
\put(125.00,15.00){\circle*{5.20}}
\put(125.00,15.00){\line(-1,0){4.67}}
\put(10.33,15.00){\makebox(0,0)[cc]{}}
\put(21.00,21.00){\makebox(0,0)[cc]{}}
\put(20.33,10.33){\makebox(0,0)[cc]{}}
\put(66.67,21.00){\makebox(0,0)[cc]{}}
\put(66.67,11.00){\makebox(0,0)[cc]{}}
\put(81.33,20.33){\makebox(0,0)[cc]{}}
\put(88.00,17.00){\makebox(0,0)[cc]{}}
\put(83.67,15.00){\makebox(0,0)[cc]{$+$}}
\put(111.00,21.67){\makebox(0,0)[cc]{}}
\put(113.00,15.00){\makebox(0,0)[cc]{$+$}}
\put(117.33,16.67){\makebox(0,0)[cc]{}}
\put(130.00,19.33){\makebox(0,0)[cc]{}}
\put(142.00,18.00){\makebox(0,0)[cc]{}}
\put(130.00,11.33){\makebox(0,0)[cc]{}}
\put(140.00,17.00){\line(-2,-1){4.00}}
\put(136.00,15.00){\line(2,-1){4.00}} \special{em:linewidth 1.2pt}
\linethickness{1.2pt} \emline{125.00}{16.67}{1}{136.00}{15.00}{2}
\emline{136.00}{15.00}{3}{125.00}{13.33}{4}
\emline{95.00}{15.00}{5}{105.00}{20.00}{6}
\emline{105.00}{20.00}{7}{105.00}{10.00}{8}
\emline{105.00}{10.00}{9}{95.00}{15.00}{10}
\emline{75.00}{19.00}{11}{63.00}{16.67}{12}
\emline{63.00}{16.67}{13}{63.00}{13.33}{14}
\emline{63.00}{13.33}{15}{75.00}{11.00}{16}
\put(50.00,15.00){\makebox(0,0)[cc]{$+$}}
\put(57.00,17.00){\makebox(0,0)[cc]{}}
\put(96.00,15.00){\circle*{3.33}}
\put(98.67,19.33){\makebox(0,0)[cc]{}}
\put(99.00,11.00){\makebox(0,0)[cc]{}}
\end{picture}

\unitlength=1.00mm \special{em:linewidth 0.4pt}
\linethickness{0.4pt}
\begin{picture}(140.00,21.00)(18,0)
\special{em:linewidth 0.4pt} \linethickness{0.4pt}
\put(23.00,15.00){\circle*{5.20}}
\put(65.00,13.00){\rule{12.00\unitlength}{4.00\unitlength}}
\put(100.00,9.00){\rule{4.00\unitlength}{12.00\unitlength}}
\put(89.00,15.00){\makebox(0,0)[cc]{$+$}}
\put(106.00,19.00){\line(-1,0){8.00}}
\put(98.00,11.00){\line(1,0){8.00}}
\put(75.00,11.00){\line(0,1){8.00}}
\put(67.00,19.00){\line(0,-1){8.00}}
\put(20.00,18.00){\line(1,-1){6.00}}
\put(26.00,18.00){\line(-1,-1){6.00}}
\put(35.00,15.00){\makebox(0,0)[cc]{$=$}}
\put(125.00,11.00){\circle*{3.33}}
\put(125.00,19.00){\circle*{3.33}}
\put(133.00,19.00){\circle*{3.33}}
\put(133.00,11.00){\circle*{3.33}}
\put(133.00,11.00){\line(1,0){3.67}}
\put(133.00,19.00){\line(1,0){3.67}}
\put(125.00,19.00){\line(-1,0){3.67}}
\put(121.33,11.00){\line(1,0){3.67}}
\put(114.00,15.00){\makebox(0,0)[cc]{$+$}}
\put(44.00,12.00){\line(1,1){6.00}}
\put(44.00,18.00){\line(1,-1){6.00}}
\put(118.00,11.00){\makebox(0,0)[cc]{}}
\put(118.00,19.00){\makebox(0,0)[cc]{}}
\put(140.00,19.00){\makebox(0,0)[cc]{}}
\put(129.00,20.67){\makebox(0,0)[cc]{}}
\put(123.00,15.67){\makebox(0,0)[cc]{}}
\put(129.00,9.00){\makebox(0,0)[cc]{}}
\put(139.00,15.33){\makebox(0,0)[cc]{}}
\put(94.67,11.00){\makebox(0,0)[cc]{}}
\put(94.67,19.00){\makebox(0,0)[cc]{}}
\put(108.67,19.00){\makebox(0,0)[cc]{}}
\put(77.33,20.00){\makebox(0,0)[cc]{}}
\put(68.67,20.33){\makebox(0,0)[cc]{}}
\put(65.67,9.67){\makebox(0,0)[cc]{}}
\put(18.00,10.33){\makebox(0,0)[cc]{}}
\put(18.00,19.00){\makebox(0,0)[cc]{}}
\put(27.00,19.00){\makebox(0,0)[cc]{}}
\put(27.33,10.33){\makebox(0,0)[cc]{}} \special{em:linewidth
1.2pt} \linethickness{1.2pt}
\emline{125.00}{19.00}{1}{133.00}{19.00}{2}
\emline{133.00}{19.00}{3}{133.00}{11.00}{4}
\emline{133.00}{11.00}{5}{125.00}{11.00}{6}
\emline{125.00}{11.00}{7}{125.00}{19.00}{8}
\put(57.00,15.00){\makebox(0,0)[cc]{$+$}}
\end{picture}

\unitlength=1.00mm \special{em:linewidth 0.4pt}
\linethickness{0.4pt}
\begin{picture}(127.00,46.67)(18,0)
\special{em:linewidth 0.4pt} \linethickness{0.4pt}
\put(48.00,15.33){\line(0,1){3.33}}
\put(117.00,17.00){\line(0,1){1.67}}
\put(117.00,15.00){\line(0,1){2.00}}
\put(48.00,17.00){\line(0,-1){1.67}}
\put(25.00,39.00){\rule{12.00\unitlength}{4.00\unitlength}}
\put(34.00,11.00){\rule{4.00\unitlength}{12.00\unitlength}}
\put(48.00,17.00){\circle*{5.20}}
\put(32.00,13.00){\line(1,0){3.00}}
\put(35.00,21.00){\line(-1,0){3.00}}
\put(48.00,18.67){\line(1,0){4.00}}
\put(48.00,15.33){\line(1,0){4.00}}
\put(27.00,45.00){\line(0,-1){8.00}}
\put(35.00,37.00){\line(0,1){8.00}}
\put(105.00,12.00){\circle*{3.33}}
\put(105.00,22.00){\circle*{3.33}}
\put(117.00,17.00){\circle*{5.20}}
\put(105.00,22.00){\line(-1,0){3.67}}
\put(101.33,12.00){\line(1,0){3.67}}
\put(117.00,18.67){\line(1,0){4.33}}
\put(117.00,15.33){\line(1,0){4.33}}
\put(68.00,11.00){\rule{4.00\unitlength}{12.00\unitlength}}
\put(70.00,21.00){\line(-1,0){4.00}}
\put(70.00,13.00){\line(-1,0){4.00}}
\put(83.00,13.00){\circle*{3.33}}
\put(83.00,21.00){\circle*{3.33}}
\put(45.00,41.00){\makebox(0,0)[cc]{$=$}}
\put(48.00,18.67){\line(0,-1){3.33}}
\put(75.00,41.00){\circle*{5.20}}
\put(75.00,42.67){\line(1,0){4.00}}
\put(75.00,39.33){\line(1,0){4.00}}
\put(113.00,45.00){\circle*{3.33}}
\put(113.00,37.00){\circle*{3.33}}
\put(113.00,37.00){\line(1,0){3.67}}
\put(113.00,45.00){\line(1,0){3.67}}
\put(89.00,41.00){\makebox(0,0)[cc]{$+$}}
\put(127.00,41.00){\makebox(0,0)[cc]{$+$}}
\put(26.00,35.00){\makebox(0,0)[cc]{}}
\put(25.33,46.67){\makebox(0,0)[cc]{}}
\put(36.33,46.67){\makebox(0,0)[cc]{}}
\put(36.33,35.00){\makebox(0,0)[cc]{}}
\put(56.33,42.67){\makebox(0,0)[cc]{}}
\put(56.33,38.67){\makebox(0,0)[cc]{}}
\put(66.67,45.00){\makebox(0,0)[cc]{}}
\put(81.67,42.67){\makebox(0,0)[cc]{}}
\put(96.00,38.67){\makebox(0,0)[cc]{}}
\put(96.00,43.33){\makebox(0,0)[cc]{}}
\put(105.00,45.67){\makebox(0,0)[cc]{}}
\put(119.33,45.00){\makebox(0,0)[cc]{}}
\put(124.67,18.67){\makebox(0,0)[cc]{}}
\put(112.00,22.33){\makebox(0,0)[cc]{}}
\put(98.00,22.00){\makebox(0,0)[cc]{}}
\put(98.00,12.00){\makebox(0,0)[cc]{}}
\put(93.00,17.00){\makebox(0,0)[cc]{$+$}}
\put(90.00,21.00){\makebox(0,0)[cc]{}}
\put(76.67,24.00){\makebox(0,0)[cc]{}}
\put(62.00,21.00){\makebox(0,0)[cc]{}}
\put(62.00,13.00){\makebox(0,0)[cc]{}}
\put(58.33,17.00){\makebox(0,0)[cc]{$+$}}
\put(54.00,18.67){\makebox(0,0)[cc]{}}
\put(43.00,22.00){\makebox(0,0)[cc]{}}
\put(28.33,21.00){\makebox(0,0)[cc]{}}
\put(28.00,13.00){\makebox(0,0)[cc]{}}
\put(102.00,41.00){\line(-2,1){4.00}}
\put(98.00,39.00){\line(2,1){4.00}}
\put(63.00,41.00){\line(-2,1){4.00}}
\put(63.00,41.00){\line(-2,-1){4.00}}
\put(83.00,13.00){\line(1,0){4.00}}
\put(83.00,21.00){\line(1,0){4.00}} \special{em:linewidth 1.2pt}
\linethickness{1.2pt} \emline{36.00}{21.00}{1}{48.00}{18.67}{2}
\emline{48.00}{15.33}{3}{36.00}{13.00}{4}
\emline{70.00}{21.00}{5}{83.00}{21.00}{6}
\emline{83.00}{21.00}{7}{83.00}{13.00}{8}
\emline{83.00}{13.00}{9}{70.00}{13.00}{10}
\emline{117.00}{18.67}{11}{105.00}{22.00}{12}
\emline{105.00}{22.00}{13}{105.00}{12.00}{14}
\emline{105.00}{12.00}{15}{117.00}{15.33}{16}
\emline{102.00}{41.00}{17}{113.00}{45.00}{18}
\emline{113.00}{45.00}{19}{113.00}{37.00}{20}
\emline{113.00}{37.00}{21}{102.00}{41.00}{22}
\emline{63.00}{41.00}{23}{75.00}{42.67}{24}
\emline{75.00}{42.67}{25}{75.00}{39.33}{26}
\emline{75.00}{39.33}{27}{63.00}{41.00}{28}
\put(112.00,12.00){\makebox(0,0)[cc]{}}
\put(77.00,11.00){\makebox(0,0)[cc]{}}
\put(43.00,12.00){\makebox(0,0)[cc]{}}
\put(68.00,38.00){\makebox(0,0)[cc]{}}
\put(106.00,37.00){\makebox(0,0)[cc]{}}
\end{picture}

\noindent(all contributions of tadpole diagrams are dropped out).
Here the thick and thin lines represent the full (within the
planar parquet approximation) and bare propagators respectively

\vspace{10mm}

\unitlength=1mm
\begin{picture}(112.00,15.00)(10,0)
\special{em:linewidth 0.4pt} \linethickness{0.4pt}
\put(90.00,15.00){\line(1,0){15.00}}
\put(47.00,15.00){\makebox(0,0)[lc]{$=~~~D~,$}}
\put(32.67,12.00){\makebox(0,0)[cc]}
\put(97.67,12.00){\makebox(0,0)[cc]}
\put(112.00,15.00){\makebox(0,0)[lc]{$=~~~1~,$}}
\special{em:linewidth 1.2pt} \linethickness{1.2pt}
\emline{25.00}{15.00}{1}{40.00}{15.00}{2}
\end{picture}

\vspace{-15mm}

$$~$$

\vspace{5mm}

\unitlength=1.00mm \special{em:linewidth 0.4pt}
\linethickness{0.4pt}
\begin{picture}(107.00,20.33)(10,0)
\put(35.00,15.00){\circle*{5.20}}
\put(35.00,15.00){\line(-1,0){5.67}}
\put(35.00,15.00){\line(1,1){4.00}}
\put(35.00,15.00){\line(1,-1){4.00}}
\put(29.00,13.00){\makebox(0,0)[cc]{$$}}
\put(40.00,17.67){\makebox(0,0)[cc]{$$}}
\put(48.00,15.00){\makebox(0,0)[lc]{$=~~~\Gamma_3~,$}}
\put(95.00,15.00){\circle*{5.20}}
\put(91.00,19.00){\line(1,-1){8.00}}
\put(99.00,19.00){\line(-1,-1){8.00}}
\put(91.67,9.67){\makebox(0,0)[cc]{$$}}
\put(92.67,20.33){\makebox(0,0)[cc]{$$}}
\put(100.00,18.00){\makebox(0,0)[cc]{$$}}
\put(107.00,15.00){\makebox(0,0)[lc]{$=~~~\Gamma _4$}}
\end{picture}

\vspace{-20mm}

$$~$$
are the full three- and four-point vertices,

\vspace{5mm}

\unitlength=1mm \special{em:linewidth 0.4pt} \linethickness{0.4pt}
\begin{picture}(110.00,22.00)(10,0)
\put(30.00,13.00){\rule{12.00\unitlength}{4.00\unitlength}}
\put(95.00,9.00){\rule{4.00\unitlength}{12.00\unitlength}}
\put(32.00,11.00){\line(0,1){8.00}}
\put(40.00,19.00){\line(0,-1){8.00}}
\put(93.00,19.00){\line(1,0){8.00}}
\put(93.00,11.00){\line(1,0){8.00}}
\put(32.00,9.00){\makebox(0,0)[cc]{}}
\put(32.00,22.00){\makebox(0,0)[cc]{}}
\put(40.00,22.00){\makebox(0,0)[cc]{}}
\put(91.00,11.00){\makebox(0,0)[cc]{}}
\put(91.00,19.00){\makebox(0,0)[cc]{}}
\put(103.00,19.00){\makebox(0,0)[cc]{}}
\put(50.00,15.00){\makebox(0,0)[lc]{$=~~~H~,$}}
\put(110.00,15.00){\makebox(0,0)[lc]{$=~~~V~$}}
\end{picture}

\vspace{-5mm}

\noindent are the parts of the four-point vertex function that are
2PR in the $t$-channel ($s$-channel) and not 2PR in the
$s$-channel ($t$-channel). The vertices $V$ and $H$ are related by
the cyclic permutation of external points.


Let us apply the foregoing considerations to the zero-dimensional
matrix Higgs model with the action
\begin{equation}
    S=\tr \left(-\frac 1 2M^2+\frac{g}{N}M^4\right).
\end{equation}
The classical vacua are
\[
    \dfrac{\delta S}{\delta M}=0\Leftrightarrow
    \left\{
    \begin{array}{l}
        M=0,
        \\
        \tr M^2=\dfrac{N^2}{4g}.
    \end{array}\right.
\]

There exist two possible ways to get the perturbative solution: we
can write down planar parquet equations in the vicinity of the
false vacuum ($M=0$) and near the true vacuum ($\tr M^2=N^2/4g$).

In the first case we have the following set of equations on $D$,
$\Gamma_4$, $H$, $V$
\begin{equation}
\left\{
    \begin{array}{l}
    D=-1-8gD^2-4gD^4\Gamma_4
\\
    \Gamma_4=-4g+H+V
\\
    H=-4gD^2\Gamma^4+VD^2\Gamma_4
\\
    V=-4gD^2\Gamma^4+HD^2\Gamma_4
    \end{array}
    \right.
\end{equation}

As one can see this is a set of four equations for four variables.
Excluding $\Gamma_4$, $V$, $H$ one get the following equation
\begin{equation}\label{Parquet_false}
    64g^3D^6+16g^2D^5+112g^2D^4+20gD^3+(1+20g)D^2+2D+1=0.
\end{equation}

It can be solved in the limit of small $g$. There exist two roots
that have no singularities as $g\to 0$
\begin{equation}\label{parquetfalse12}
\left\{
    \begin{array}{l}
    D^{(1)}=-1-12g+o(g),
    \\
    D^{(2)}=-1-8g+o(g).
    \end{array}\right.
\end{equation}

Moreover, there exists a solution $D^{(3)}$ behaving like
$\dfrac{\alpha}{g}$ as $g\to 0$. One can write down
the following equation for $\alpha$
\begin{equation}
    64\alpha^6+16\alpha^5=0,
\end{equation}
wherefrom $\alpha=-\dfrac 1 4$ and
\begin{equation}\label{parquetfalse3}
    D^{(3)}=-\dfrac{1}{4g}+o(g).
\end{equation}

The following table contains results of numerical calculations of
(\ref{Parquet_false}). The solutions exist if $g<0,038$.

\vspace{5mm}

{\footnotesize\noindent\begin{tabular}{|c|c|c|c|c|c|c|c|}
  \hline
  $g$         & 0   & $10^{-6}$  &$10^{-5}$  &$10^{-4}$& $10^{-3}$  &$10^{-2}$   &$10^{-1}$\\
  \hline
  $D^{(1)}$   & $-$1& $-$1,000012&$-$1,000120&$-$1,001203& $-$1,012331& $-$1,169973& --- \\
  $D^{(2)}$   & $-$1& $-$1,000008&$-$1,000080&$-$1,000801& $-$1,008113& $-$1,093576& --- \\
  $D^{(3)}$   & --- & $-$249997,0& $-$24996,9&
  $-$2496,9& $-$246,9   &$-$21,8     &---\\
  \hline
\end{tabular}
}
\vspace{10mm}

The second case is more delicate. Consider a shift to the true
vacuum
\begin{equation}\label{shift}
    M_{\alpha\beta}=R_{\alpha\beta}+\sqrt{\dfrac{N}{4g}}I_{\alpha\beta},\quad
    \tr R=0.
\end{equation}
Hence, we get the following action
\begin{equation}\label{Higgs_shifted}
    S=\tr\left(R^2+
    \frac{2\sqrt g}{\sqrt N}R^3+\frac g N R^4-\frac{N^2}{16
    g}\right).
\end{equation}
For convenience one can rescale the fields
$R=\dfrac{Q}{\sqrt 2}$
\[
    S=\tr\left(\frac{Q^2}{2}+
    \frac{\sqrt g}{\sqrt 2\sqrt{N}}Q^3+\frac{g}{4N} Q^4-\frac{N^2}{16
    g}\right).
\]
It contains both cubic and quartic interaction terms.

Thus, in the zero-dimensional case the planar parquet equations on
$\widetilde D$, $\widetilde \Gamma_3$,
$\widetilde \Gamma_4$\footnote{Here ${\widetilde D=\,<\tr Q^2>}$, ${\widetilde \Gamma_3=\,<\tr Q^3>}$,
${\widetilde \Gamma_4=\,<\tr Q^4>}$ are Green functions
full within the planar parquet approximation.}
look like
\begin{equation}\label{parquet}
    \left\{
    \begin{array}{l}
        \widetilde D=1-\dfrac{3\sqrt 2}{2}\;\sqrt{g} \widetilde D^3\widetilde \Gamma_3-2g\widetilde D^2-g\widetilde D^4\widetilde \Gamma_4
        \\
        \widetilde \Gamma_3=-\dfrac{3\sqrt 2}{2}\;\sqrt{g}+\widetilde D^2\widetilde V\widetilde \Gamma_3+\widetilde D^3\widetilde \Gamma_3^3-
        g\widetilde \Gamma_3\widetilde D^2
        \\
        \widetilde \Gamma_4=-g+\widetilde H+\widetilde V+\widetilde D^4\widetilde \Gamma_3^4
        \\
        \widetilde H=\widetilde D^2\widetilde \Gamma_4\widetilde V+\widetilde D^3\widetilde \Gamma_3^2\widetilde V+
        \widetilde D^3\widetilde \Gamma_4\widetilde \Gamma_3^2-g\widetilde D^2\widetilde \Gamma_4-
        g\widetilde D^3\widetilde \Gamma^2_3
        \\
        \widetilde V=\widetilde D^2\widetilde \Gamma_4\widetilde H+\widetilde D^3\widetilde \Gamma_3^2\widetilde H+\widetilde D^3
        \widetilde \Gamma_4\widetilde \Gamma_3^2-g\widetilde D^2\widetilde \Gamma_4-
        g\widetilde D^3\widetilde \Gamma^2_3
    \end{array}
    \right.
\end{equation}

This set can be solved as $g\to 0$
\begin{equation}
    \widetilde D=1+\dfrac 5 2 g+o(g).
\end{equation}
This set of equations gives a sensible physical solution if
$g<0,037$.

\vspace{5mm}

{\small\noindent\begin{tabular}{|c|c|c|c|c|c|c|c|}
\hline
  $g$            & 0& $10^{-6}$&$10^{-5}$&$10^{-4}$&$10^{-3}$&$10^{-2}$&$10^{-1}$
  \\
  \hline
  $\widetilde D$ & 1& 1,000002 &1,000060 &1,000250 &1,002504 & 1,030238&-2,220680\\
  \hline
\end{tabular}}

\vspace{5mm}

Hence, the true value of the propagator is
\begin{equation}\label{True}
    D=\,<\tr M^2>=\frac{1}{4g}+\frac 1 2\;<\tr Q^2>=\frac{1}{4g}+\frac 1 2+\frac 5 4 g+o(g).
\end{equation}

\section{Planar approximation}

In this section we study the zero-dimensional matrix Higgs model by means of the
approach proposed in~\cite{BIPZ}. A specific feature of this model is that it
contains two-cut solutions parameterized by a special parameter \cite{Shimamune, Ar84,Cicuta} .

In \cite{BIPZ} one considered the large $N$ limit in the model
\begin{equation}\label{plusmodel}
    S=\tr\left(\frac 1 2M^2+\frac g N M^4\right).
\end{equation}
By means of the steepest descent method the vacuum energy
\begin{equation}
    e^{-N^2 E_0(g)}=\lim_{N\to\infty}\int DM \;e^{-S}
\end{equation}
and Green functions
\begin{equation}
    G_n(x_1,...,x_n)=\lim_{N \to
    \infty}\frac{1}{N^{1+n/2}}\dfrac{\int DM \tr( M(x_1)... M(x_n))\exp
    (-S)}{\int DM \exp (-S)}
\end{equation}
were calculated.

The results are presented below
\begin{equation}
    E_0(g)=\int_{X}d\lambda\, u(\lambda)\left(\frac{1}{2}\lambda^2+
    g\lambda^4\right)-\iint_{X}d\mu d\lambda u(\lambda) u(\mu)
    \ln|\lambda-\mu|,
\end{equation}
\begin{equation}
    G_{2p}=\int_X \lambda^{2p}u(\lambda)d\lambda,
\end{equation}
here one introduced the eigenvalue density function $u(\lambda)$
such that $${\int_X u(\lambda)d\lambda=1},$$ where $X$ is the support
of $u(\lambda)$. This density function is the solution of the
following singular equation
\[
    \frac 1 2\lambda+2g\lambda^3=\;v.p.\;\int_{X}\frac{u(\mu)}{\lambda-\mu}d\mu.
\]
The model (\ref{plusmodel}) admits so called one-cut solutions,
i.e. when $X$ has the form of a single segment $(2a,\, 2b)$. This
support is uniquely defined by a set of algebraic equations.

The matrix Higgs model
\begin{equation}\label{Higgsmodel}
    S=\tr\left(-\frac 1 2M^2+\frac g N M^4\right).
\end{equation}
is interesting because it is the simplest model where exist both
one-cut and multi-cut (two-cut) solutions. As it was mentioned in
\cite{Ar84, Cicuta} in the case of two-cut solution there exists
certain freedom, to fix it one has to introduce an extra
parameter. It is associated with the order parameter which governs
the phase structure of the system.

Consider the one-cut solution, i.e. let $X=(2a,2b)\equiv
(t-s,t+s)$. There exist two solutions. The first one corresponds
to symmetric support $t=0$. It gives the density function
\begin{equation}
    u(\lambda)=\frac{1}{2\pi}\left(8a^2g+4g\lambda^2-1\right)\sqrt{(2a)^2-\lambda^2},\qquad
    g>0,
\end{equation}
where $a$ is subject to
\begin{equation}
    12ga^4-a^2-1=0,
\end{equation}
and the two-point Green function
\begin{equation}\label{planar1sym}
    G_2=\int_{-2a}^{2b} d\lambda \;\lambda^2
    u(\lambda)=\frac{a^2(4+a^2)}{3}=\frac{1}{432g^2}+\frac{1}{6g}+1-8g+o(g).
\end{equation}

The second solution corresponds to the non-symmetric support
\begin{equation}
    t^2=\frac{3+2\sqrt{1-60g}}{20g},\qquad
    s^2=\frac{1-\sqrt{1-60g}}{15g},\qquad 0<g\leq1/60.
\end{equation}
The density function looks like
\begin{equation}
     u(\lambda)=\frac{1}{2\pi}\left(4g\lambda^2+4gt\lambda+4gt^2+2gs^2-1\right)
    \sqrt{2t\lambda-\lambda^2+s^2-t^2}
\end{equation}
and the two-point Green function is
\begin{equation}\label{planar1}
    G_2=-\frac{s^2t^2}{4}-\frac{s^4}{16}+3gs^2t^4+3gs^4t^2+\frac{gs^6}{4}=\frac{1}{4g}-1-10g+o(g).
\end{equation}

As in the previous section one can perform these calculations in
the different way: one can make the shift (\ref{shift}) to the
true vacuum and apply the same technique to the action
(\ref{Higgs_shifted}). But this case can be solved only in the
limit of small $g$. The results are presented below:

\noindent the support $X=(2a,2b)$ is
\begin{equation}
    \left\{
    \begin{array}{l}
    a=-\dfrac{1}{\sqrt 2}-\dfrac 3
    2\sqrt{g}-\dfrac{15}{4}\sqrt{2}g+o(g),
    \\
    b=\dfrac{1}{\sqrt 2}-\dfrac 3
    2\sqrt{g}+\dfrac{15}{4}\sqrt{2}g+o(g),
    \end{array}
    \right.
\end{equation}
the density function is
\begin{multline}
    u(\lambda)=\frac 1\pi(2g\lambda^2+2g(a+b)\lambda+3\sqrt{g}\lambda+2g(a+b)^2+3\sqrt{g}(a+b)+g(a-b)^2+1)
    \times
    \\
    \times\sqrt{(2a-\lambda)(\lambda-2b)},
\end{multline}
and the behavior of the two-point Green function as $g\to 0$, taking into
account (\ref{True}), is
\begin{equation}\label{planar2}
    G_2=\frac{1}{4g}+\frac 1 2+8g+o(g).
\end{equation}

As one can observe two different values for $G_2$ (\ref{planar1})
and (\ref{planar2}) are not the same, they show similar behavior
as $g\to 0$.

Consider the two-cut solution. The simplest case is the symmetric
support ${X=(-2b,-2a)\cup(2a,2b)}$. By means of the previous
procedure one can get the following:

\noindent the support $X$ is
\begin{equation}
    a^2=\frac{1-4\sqrt g}{4g},\qquad b^2=\frac{1+4\sqrt g}{4g},
    \quad 0<g\leq 1/16,
\end{equation}
the density function is
\begin{equation}
    u(\lambda)=\frac{|\lambda|}{2\pi}\sqrt{8g\lambda^2-16g^2\lambda^4-1+16g},
\end{equation}
and the two-point Green function
\begin{equation}\label{planar3}
    G_2=\frac{1}{4g}.
\end{equation}
One can see that in this case the final result is exact.

One has to notice that (\ref{planar1}), (\ref{planar2}) and (\ref{planar3})
have the same behavior as $g\to 0$.

\section{Parquet planar generating functional}

1. \textit{Planar generating functional}

For the sake of simplicity we will study the model with the
following action
\begin{equation}\label{action}
    S=\frac{1}{2}\mbox{Tr }M^2+\frac{g}{4N}\mbox{Tr }M^4.
\end{equation}

The planar Schwinger--Dyson equations in this case are
\begin{equation}\label{SD}
    \Pi_n+g\Pi_{n+2}=\sum_{i=0}^{n-2}\Pi_i\Pi_{n-i-2}, \qquad n>2.
\end{equation}

To solve them one introduces the following functional
\[
    F(x)=\sum_{n=0}^{\infty}x^n\Pi_n.
\]
where $\Pi_0=1$ and $\Pi_{2k+1}=0$ since the measure in
(\ref{Greens}) is invariant under $M\rightarrow - M$.

It can be easily seen that $F(x)$ satisfies
\begin{equation}
    x^4F^2-(g+x^2)F+g+x^2+gx^2\Pi_2=0.
\end{equation}
Hence, the generating functional has the form
\begin{equation}
    F=\dfrac{x^2+g-\sqrt{(x^2+g)^2-4x^4(g+x^2+gx^2\Pi_2)}}{2x^4}.
\end{equation}

Therefore, the Green functions $\Pi_n$ are expressed in terms of
$\Pi_2$, i.e. to know the Green series it is necessary to write
down an equation for $\Pi_2$. This fact was discovered in
\cite{Cv}. One must mention that the approach based on the planar
Scwinger--Dyson equations is not self-sufficient: it does not give
such an equation on $\Pi_2$. One can write down the required
equation within the approach proposed in \cite{BIPZ}. For the
action (\ref{action}) this equation looks like
\begin{equation}
    27g^2\Pi_2^2+(1+18g)\Pi_2-1-16g=0.
\end{equation}

2. \textit{Planar parquet generating functional}

As it was said in section \ref{section1}, the planar parquet
approximation takes into account only a subset of all planar
diagrams and is defined as an approximative solution of the
Schwinger--Dyson equations. In this limit the Schwinger--Dyson
equations are reduced to a certain set of equations on the "basic"
Green functions, i.e. $\Pi_2$, $\Gamma_3$, $\Gamma_4$. For the
action (\ref{action}) this set looks like
\begin{equation}
    \left\{\begin{array}{l}
    \Pi_2=1-2g\Pi_2^2-g\Pi_2^4\Gamma_4
    \\
    \Gamma_4=-g+H+V
    \\
    H=-g\Pi_2^2\Gamma_4+V\Pi_2^2\Gamma_4
    \\
    V=-g\Pi_2^2\Gamma_4+H\Pi_2^2\Gamma_4
    \end{array}\right.
\end{equation}
or
\begin{equation}{\label{S}}
    \left\{\begin{array}{l}
    \Pi_2=1-2g\Pi_2^2-g\Pi_2^4\Gamma_4
    \\
    \Gamma_4=-g+\dfrac{2g\Pi_2^2\Gamma_4}{\Pi_2^2\Gamma_4-1}
    \end{array}\right.
\end{equation}

The higher Green functions are constructed in terms of $\Pi_2$ and
$\Gamma_4$.

The first equation is nothing else but the Schwinger--Dyson
{equation (\ref{SD})} for $n=2$. It can be easily seen since the
full 4-point function $\Pi_4$ in the planar parquet limit is
\[
    \Pi_4=2\Pi_2^2+\Pi_2^4\Gamma_4,
\]
so
\[
    \Pi_2=1-g\Pi_4.
\]

Hence, the following equation on $\Pi_2$ derived from (\ref{S})
\begin{equation}
    g^3\Pi_2^6+g^2\Pi_2^5+5g^2\Pi_2^4+5g\Pi_2^3+(1-5g)\Pi_2^2-2\Pi_2+1=0
\end{equation}
is the required equation on $\Pi_2$.

3. \textit{Parquet Higgs model}

Consider the following action
\begin{equation}\label{action1}
    S=\frac{1}{2}\mbox{Tr }M^2+\frac{\lambda}{3\sqrt{N}}\mbox{Tr }M^3+\frac{g}{4N}\mbox{Tr }M^4.
\end{equation}

The Schwinger--Dyson equations in this case are
\begin{equation}
    \Pi_n+\lambda \Pi_{n+1}+g\Pi_{n+2}=\sum_{i=0}^{n-2}\Pi_i\Pi_{n-i-2}, \qquad n>2.
\end{equation}
The functional and its equation are
\begin{equation}\label{gen_fun}
    F(x)=\sum_{n=0}^\infty x^n \Pi_n
\end{equation}
\begin{multline}
    x^4F^2-(x^2+\lambda x+g)F+x^2+\lambda x+g+\\
    \Pi_1 x(x^2+\lambda x+g)+\Pi_2x^2(\lambda x+g)+\Pi_3gx^3=0
\end{multline}
Thus, the generating functional can be calculated in terms of 3 arbitrary
constants $\Pi_1$, $\Pi_2$, $\Pi_3$.

The Higgs model in the vicinity of the true vacuum is the same as
(\ref{action1}) with $\lambda=\lambda(g)$.

The planar parquet set of equations is given in Section
\ref{section1}. It can be easily seen that the planar parquet
approximation gives only two equations. To get the system closed
one proposes the following equation on tadpole diagrams
\begin{equation}
    \Pi_1=\lambda\Pi_3.
\end{equation}

Hence, the planar parquet approximation gives the exact solution
in terms of the generating functional (\ref{gen_fun}) together
with the set of equations on $\Pi_1$, $\Pi_2$, $\Pi_3$.

In \cite{tH1} the action was considered in the planar limit by
means of the generating functional technique. It was said that
the arbitrary parameters (boundary conditions) can be derived by
careful studying the holomorphic properties of the generating
functional.

\section{Conclusion}

We have considered two different approaches to the matrix Higgs
model. We have got the following results.

1. The planar two-point functions for
the symmetric one-cut (\ref{planar1sym}), non-symmetric one-cut cases
(\ref{planar1}) and (\ref{planar2}) have the following asymptotics as $g\to\infty$
\begin{equation*}
    \begin{array}{l}
    G^{(1)}_2=\dfrac{1}{432g^2}+\dfrac{1}{6g}+1-8g+o(g),
    \\
    G^{(2)}_3=\dfrac{1}{4g}-1-10g+o(g),
    \\
    G^{(3)}_2=\dfrac{1}{4g}+\dfrac 1 2+8g+o(g),
    \end{array}
\end{equation*}
and in the two-cut case the planar two-point function (\ref{planar3}) is
\[
    G_2^{(4)}=\frac{1}{4g}.
\]

2. The asymptotical expressions for the parquet two-point
functions in the false vacuum (\ref{parquetfalse12}),
(\ref{parquetfalse3}) are
\begin{equation*}
    \begin{array}{l}
    D^{(1)}_2=-1-12g+o(g),
    \\
    D^{(2)}_2=-1-8g+o(g),
    \\
    D^{(3)}_2=-\dfrac{1}{4g}+o(g),
    \end{array}
\end{equation*}
and in the true vacuum the two-point function has the following asymptotic (\ref{True})
\[
    D^{(4)}_2=\dfrac{1}{4g}+\dfrac 1 2+\dfrac 5 4 g+o(g).
\]

Hence, $D^{(4)}_2$, two-point Green function computed within the
planar parquet approach in the true vacuum, coincides with
$G^{(3)}_2$, two-point Green function computed within the planar
approximation also in the true vacuum.

As it was mentioned the parquet planar approach gives a very good
agreement with the exact results in the one-cut case for the case
of the positive mass square. In the case of the matrix Higgs
model the situation seems more delicate because of the its
multi-phase structure. Nevertheless, the planar parquet
approximation leads to a rather good agreement at least in the
small coupling limit.

We have shown that in the planar parquet approximation it is
possible to construct the generating functional for the Green
functions. Besides, we have shown that the generating functional
in the planar limit depends on the approximation chosen and,
therefore, can be restricted to a subset of diagrams just by
modifying initial conditions of the system.

$$~$$
{\bf Acknowledgments}

We would like to thank I.Ya.~Aref'eva for useful
discussions. This work is supported by RFBR grant 02-01-00695 and
RFBR grant for leading scientific schools.
$$~$$



\begin{thebibliography}{99}

\bibitem{LAK} L.D. Landau,
A.A.~Abrikosov  and I.M.~Khalatnikov, Dokl. Acad.  Nauk, 95
(1954) 497, 773, 1177;

L.D.~Landau, in: Niels Bohr and the Development of Physics, Ed.
W.Pauli. --- New York: Mc Grow-Hill, 1955;

\bibitem{TM}I.T. Dyatlov, V.V. Sudakov, K.A.
Ter-Martirosyan, \textit{Asymptotic meson--meson dispersion
theory}, Sov. Phys. JETP, 4 (1957) 767;

A.A.~Abrikosov, A.D.~Galanin, L.P.~Gorkov, L.D.~Landau,
I.Ya.~Po\-me\-ran\-chuk and K.A.~Ter-Mar\-ti\-ro\-syan,
\textit{Possibility of formulation of a theory of strongly
interacting fermions}, Phys. Rev., 111 (1958) 321;

K.A.~Ter-Mar\-ti\-ro\-syan, \textit{Equation for vertex part
corresponding to fermion---fermion scattering}, Phys. Rev., 111
(1958) 948;

Yu.M.~Makeenko, K.A.~Ter-Mar\-ti\-ros\-yan and
A.B.~Za\-mo\-lod\-chi\-kov, \textit{On The Theory Of The Direct
Four-Fermion Interaction}, Sov. Phys. JETP, 44 (1976) 11;

\bibitem{tH} G.'t Hooft, \textit{A planar diagram theory strong interactions},
Nucl.Phys. B, 72 (1974) 461;

\bibitem{BIPZ} E. Br\'ezin, C. Itzykson, G. Parisi, J.B. Zuber, \textit{Planar diagrams},
Commun. Math. Phys., 59 (1978) 35;

\bibitem{Witten} E. Witten, \textit{Baryons in the $1/N$ expansion},
Nucl.Phys. B, 160 (1979) 57;

\bibitem{Ar96} I.Ya. Aref'eva, A.P. Zubarev, \textit{Parquet approximation
in large $N$ matrix theories}, Phys. Lett. B, 386 (1996) 258;

\bibitem{Shimamune} Y. Shimamune, \textit{On the phase structure of large $N$
matrix models and gauge models}, Phys. Lett. B, 108 (1982) 407;

\bibitem{Ar84} I.Ya. Aref'eva, A.S. Ilchev, V.K. Mitrjushkin,
\textit{Phase structure of the matrix Goldstone model in the
large N limit}, in Proc. "III International Symposium on Selected
Topics in Statistical Mechanics", Dubna, 1984, vol.1, p.20;

\bibitem{Cicuta} G.M. Cicuta, L. Molinari, E.Montaldi, \textit{Large-$N$ spontaneous magnetisation
in zero dimension}, J.Phys.A: Math. Gen., 20 (1987) L67;

\bibitem{Haan}  O.Haan, \textit{On the structure of planar field theory},
Z. Physik, C6 (1980) 345;

\bibitem{Ar81}  I.Ya. Aref'eva, \textit{Functional equations for planar graphs}, Phys. Lett. B, 104(1981)
453;

\bibitem{tH1}  G.'t Hooft, \textit{Counting planar diagrams with various restrictions},
Nucl.Phys. B, 538 (1999) 389, hep-th/9808113;

\bibitem{Cv}  P. Cvitanovi\'{c}, P.G. Lauwers, P.N. Scharbach,
\textit{The planar sector of field theories}, Nucl.Phys. B, 203
(1982) 385.

\end{thebibliography}
\end{document}